\documentclass[%
 aip,
 amsmath,amssymb,
 reprint,%
]{revtex4-1}

\usepackage{graphicx}% Include figure files
\usepackage{dcolumn}% Align table columns on decimal point
\usepackage{bm}% bold math
\usepackage[utf8]{inputenc}
\usepackage[T1]{fontenc}
\usepackage{mathptmx}
\usepackage{commath}

\begin{document}

\preprint{AIP/123-QED}

\title[]{Machine learning depinning of dislocation pileups}
% Force line breaks with \\

\author{Mika Sarvilahti}
\author{Audun Skaugen}
\author{Lasse Laurson}
\email{lasse.laurson@tuni.fi}
\affiliation{ 
$^1$Computational Physics Laboratory, Tampere University, \\
P.O. Box 692, FI-33014 Tampere, Finland%\\This line break forced with \textbackslash\textbackslash
}%

\date{\today}% It is always \today, today,

\begin{abstract}
We study a one-dimensional model of a dislocation pileup driven by an external stress and interacting
with random quenched disorder, focusing on predictability of the plastic deformation process. Upon quasistatically ramping up the
externally applied stress from zero the system responds by exhibiting
an irregular stress--strain curve consisting of a sequence of strain bursts,
i.e., critical-like dislocation avalanches. The strain bursts are power-law distributed
up to a cutoff scale which increases with the stress level up to a
critical flow stress value. There, the system undergoes a depinning phase
transition and the dislocations start moving indefinitely, i.e., the
strain burst size diverges. Using sample-specific information about
the pinning landscape as well as the initial dislocation configuration as
input, we employ predictive models such as linear regression, simple neural
networks and convolutional neural networks to study the predictability of
the simulated stress--strain curves of individual samples. 
Our results show that the response of the system -- including the flow stress value -- can be predicted quite well, with the correlation coefficient between predicted and actual
stress exhibiting a non-monotonic dependence on strain. 
We also discuss our attempts to predict the individual strain bursts.
\end{abstract}

\maketitle

%\begin{quotation}
%The ``lead paragraph'' is encapsulated with the \LaTeX\ 
%\verb+quotation+ environment and is formatted as a single paragraph before %the first section heading. 
%(The \verb+quotation+ environment reverts to its usual meaning after the %first sectioning command.) 
%Note that numbered references are allowed in the lead paragraph.
%%
%The lead paragraph will only be found in an article being prepared for the %journal \textit{Chaos}.
%\end{quotation}

\section{Introduction}

One of the key ideas of many disciplines including materials science
in particular is that structure and properties of materials tend to be
related. \cite{cheng2011atomic,delince2007structure}
For instance, much of metallurgy is about tuning materials
microstructure to design materials with desired mechanical properties.
\cite{delince2007structure} Typically the relation between, say, the precipitate content of a
precipitation-hardened alloy \cite{ardell1985precipitation}
and its yield strength is understood
as an average property, and indeed in the case of macroscopic
bulk samples the sample-to-sample variations in the yield stress of
identically prepared samples tend to be small. However, this changes when
dealing with samples with sizes down to the micrometer range and below:
Recent micropillar compression experiments have revealed the fluctuating,
irregular nature of small-scale crystal plasticity, \cite{uchic2004sample,
dimiduk2006scale} originating from
the collective, critical-like dynamics of interacting dislocations mediating
the deformation process. The fact that plasticity of
small crystals proceeds via an irregular sequence of strain bursts with
a broad size distribution implies also that sample-to-sample variations
of the plastic response or the stress--strain curves might be considerable, \cite{dimiduk2005size}
even if the samples have been prepared and the deformation experiments
performed using the same protocol.

This raises the important general question of what can be said
about the relation between initial structure and mechanical properties for individual samples of small crystals,
when random variations in the initial
(micro)structure become important. One way of framing the issue is to
consider what we refer to as {\it deformation predictability}: Given a small
crystalline sample with a specific arrangement of pre-existing dislocations
and possibly of some other defects interfering with dislocation motion,
how well can one use such information to predict the plastic response of
that sample? A key challenge here is the high dimensionality of any reasonable description of the disordered initial state, combined with the possibility of non-linearities in the mapping from the initial state properties to the ensuing response. This likely implies that simple empirical laws relating sample properties to its plastic response may be difficult to formulate. Challenges of this type are a key factor behind the recent emergence of machine learning (ML) algorithms as an important part of the toolbox of scientists in a wide range of fields including also physics and materials science. \cite{zdeborova2017machine} ML has been demonstrated to be an efficient approach to address a wide spectrum of problems including materials property prediction, discovery of novel materials, etc. \cite{papanikolaou2019spatial,
raissi2018hidden,baldi2014searching,butler2018machine,raccuglia2016machine,pilania2013accelerating,liu2020material} These developments have lead to the emergence of a novel research field referred to as {\it materials informatics}, \cite{ramprasad2017machine} where informatics methods -- including ML -- are used to determine material properties that are hard to measure or compute using traditional methods. 

In general, many supervised ML algorithms including neural networks are capable of learning non-linear mappings from a high-dimensional feature vector to a desired output, and hence these methods should be useful tools when addressing the question of deformation predictability. An important opening in this direction was recently achieved by Salmenjoki {\it et al.},
\cite{salmenjoki2018machine} who applied ML to study
deformation predictability in the case of simple two-dimensional discrete
dislocation dynamics (DDD) simulations. The key idea of Ref.
\cite{salmenjoki2018machine} was that exploiting ML algorithms provides a
useful set of methods to quantify predictability of complex systems such as plastically deforming crystals. This was achieved by training ML models including neural networks and support vector machines to infer the mapping from the initial dislocation microstructure to the response of the system to applied stresses, characterized by the stress--strain curve. Following Ref. \cite{salmenjoki2018machine}, ML has been further applied to predicting stress--strain curves of plastically deforming crystals, \cite{yang2020learning} and learning the interaction kernel of dislocations. \cite{salmenjoki2018mimicking} Recently ML has also been applied to the closely related problems of, e.g., predicting the local yielding dynamics of dry foams, \cite{viitanen2020machine} the creep failure time of disordered materials, \cite{biswas2020prediction} as well as the occurrence times of "laboratory earthquakes". \cite{rouet2017machine} 

By its nature, the problem of predicting the plastic deformation process of
crystalline samples depends on details such as whether the crystal only contains pre-existing glissile dislocations
(this was the case in Ref. \cite{salmenjoki2018machine}), or if other
defect populations interacting with the dislocations are present. The latter
could include solute atoms, precipitates, voids, etc., or even grain
boundaries in the case of polycrystals. If present, a description of
these additional defects needs to be included in the
initial state of the system used as input for
the predictive ML models. The presence of such static defects within
the crystal may also change the response
of the crystal to applied stresses and thus the process to be predicted in a fundamental manner. It has been shown that the deformation dynamics of ``pure'' dislocation systems are governed by dislocation jamming, \cite{miguel2002dislocation,laurson2010dynamical} resulting in glassy dislocation dynamics characterized by ``extended criticality'', with
the cutoff scale of the size distribution of dislocation avalanches diverging with the system size at any stress level. \cite{ispanovity2014avalanches,lehtinen2016glassy,ovaska2017excitation} This can be contrasted with systems where significant
pinning of dislocations due to other defects within the crystal may instead
induce a depinning phase transition of the dislocation assembly, resulting
in critical-like dislocation dynamics only in the immediate proximity of the critical point (stress) of the
depinning transition. \cite{ovaska2015quenched,salmenjoki2020precipitate}

In this paper, we study deformation predictability in a one-dimensional
periodic model of a dislocation pileup, interacting with a quenched (frozen) random
pinning landscape. The model is perhaps the simplest possible system including
both interacting mobile dislocations and a quenched random pinning field interfering with
dislocation motion, and therefore it serves
as a useful playground to explore the ideas of ML-based deformation
predictability discussed above. It is known to exhibit a depinning phase transition at a critical flow stress value $\sigma_\textrm{flow}$, separating pinned and moving phases of the dislocation system. \cite{moretti2004depinning} We generate a large database of
stress--strain curves, each corresponding to a unique randomly generated
initial microstructure, by simulating the model with a quasistatically
increasing applied stress for different system sizes (or different numbers of dislocations, $N$). First, the statistical properties 
of the stress--strain curves and the strain bursts are analyzed, followed by training and testing of various predictive models ranging from linear regression to convolutional neural networks (CNNs) to establish mappings from the initial random microstructure (defined by both the static pinning landscape and the initial configuration of the dislocations) to the response of the system to applied stresses (i.e., the stress--strain curve).

Our results show that the different predictive models employed
are capable of learning the relation between the input and the stress--strain curves quite well, as measured by the correlation coefficient between the predicted and simulated stress value at a given strain. In particular, the sample flow stress (i.e., the sample-dependent finite size critical point of the depinning transition) can be predicted surprisingly well, with the correlation coefficient reaching values as high as 0.89 for a
regularized CNN. We also explore the predictability of individual strain bursts taking place during the deformation process. Critical avalanches are expected to be inherently unpredictable, and indeed we find approximately zero predictability for most of the strain bursts belonging to
the power law scaling regime of their size distribution. Interestingly, we also find that strain bursts belonging to the cutoffs of the distributions as well as those taking place very close to the flow stress exhibit non-vanishing predictability. 

\section{The model: edge dislocation pileup with quenched disorder}

The dislocation pileup model we study here is similar to the one in Refs. \cite{moretti2004depinning,
leoni2009slip}. It describes a system of $N$ straight, parallel edge
dislocations subject to an external shear stress $\sigma$ gliding along the
direction set by their Burgers vector (here, the $x$ direction) within
a given glide plane. By neglecting roughness of the dislocation lines,
the system may be described by a set of point dislocations with coordinates
$x_i$ representing cross sections of the dislocation lines moving in one
dimension. The dislocations interact repulsively with each other via
long-range stress fields with the stress field magnitude inversely
proportional to their mutual distance. \cite{moretti2004depinning}
In addition, the dislocations
are taken to interact via short-range forces with a set of $N_p$ randomly
positioned Gaussian pinning centres, playing here the role of quenched
disorder. In real crystals these pinning centres could be solute atoms, or immobile forest dislocations threading the plane of the pileup. The overdamped equations of motion of the $N$ point dislocations thus read
\begin{equation}
\chi \frac{\mathrm{d}x_i}{\mathrm{d}t} = \mu b^2 \sum_{j \neq i}
\frac{1}{x_i-x_j} + b\sigma + F(x_i),
\end{equation}
where $\chi$ is the effective viscosity, $x_i$ the position of the
$i$th dislocation, $\mu$ the shear modulus, $b$ the Burgers vector
magnitude, and $F$ is the pinning force landscape. We choose a pinning force consisting of Gaussian potential wells centered at pinning sites $\set{x_p}$, given by
\begin{equation}
\label{eq:pf}
F(x) = -\od{E}{x},\quad
E(x) = -\sum_{p=1}^{N_p} E_p\mathrm{e}^{-\frac{1}{2}\left(\frac{x-x_p}{s_p}\right)^2},
\end{equation}
where $x_p$ is the position of the $p$th pinning point, $E_p$ a pinning
energy scale, and $s_p$ the standard deviation of the Gaussian. For simplicity, we set
$\chi=\mu=b=1$, and $E_p = s_p^2 = 0.5$. Periodic
boundary conditions within a system of size $L$ are employed by performing
an infinite sum over the stress fields of the periodic images of each
dislocation, \cite{moretti2004depinning}
\begin{equation}
\sum_{k=-\infty}^{\infty} \frac{1}{x+kL} = \frac{\pi}{L\tan (\pi x/L)}.
\end{equation}
The average spacing between dislocations is set to $L/N=16$ and that between
pinning sites to $L/N_p=2$. The sum over pinning sites is truncated so that
only pinning sites that are closer than a cut-off distance of 8 are
included in the sum.

\begin{figure}[t!]
    \begin{center}
  \includegraphics[width=\columnwidth]{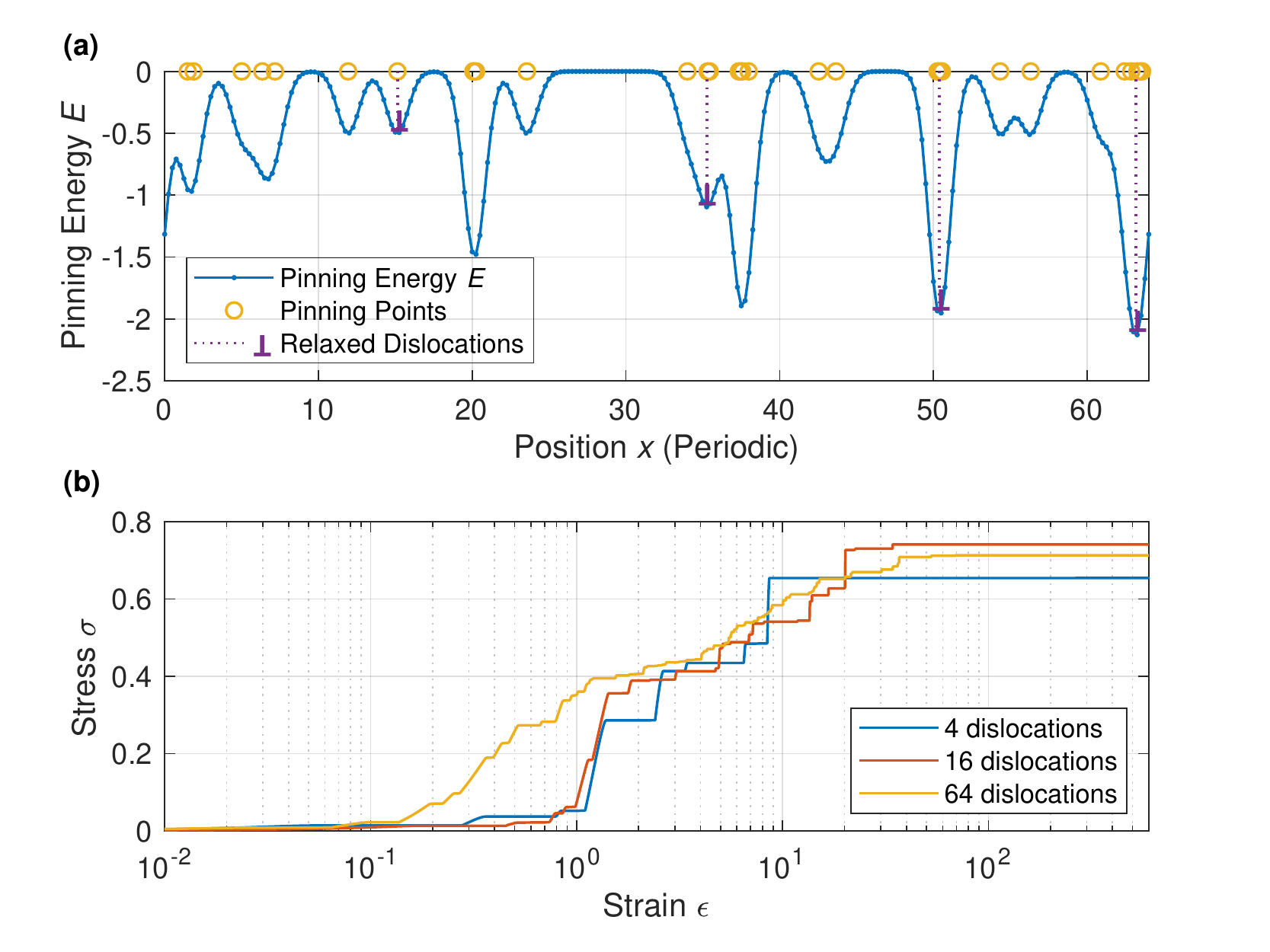}
  \caption{(a) An example of a relaxed dislocation configuration in the
    simulation model with $N=4$ (and hence of linear size $L=16N=64$). The
    randomly positioned pinning sites (there are $N_p = L/2 = 32$ of them)
    are shown as circles, and the resulting pinning energy landscape $E(x)$
    is shown with the solid line. (b) Example stress--strain curves
    $\sigma (\epsilon)$ for three different system sizes ($N=4$, 16 and 64),
    illustrating the characteristic irregular staircase shape of the curves, as
    well as the size effect where smaller systems exhibit fewer but larger strain
    bursts.}
  \label{fig:1}
\end{center}
\end{figure}

The simulations are carried out by first choosing the positions of the $N_p$ pinning sites at random from a uniform distribution
along the one-dimensional system of length $L$. $N$ dislocations
are then placed in the system with the constant spacing of $L/N=16$; this equally
spaced dislocation configuration would be the minimum energy state of the
pileup in the absence of pinning sites. The dislocations are initially allowed to
relax to a metastable configuration with $\sigma = 0$; see Fig.~\ref{fig:1}(a) for an example of a relaxed
configuration with $N=4$, showing also the pinning energy landscape. Then
$\sigma$ is increased quasistatically from zero with the rate
\begin{equation}
\frac{\mathrm{d}\sigma}{\mathrm{d}t} = \frac{m_{\sigma}}
{\mathrm{e}^{\frac{r[\bar{v}(t)-v_{\mathrm{th}}]}{v_{\mathrm{th}}}}+1},
\end{equation}
where $m_{\sigma} = 10^{-4}$ is the maximum stress rate,
$v_{\mathrm{th}}=2\cdot 10^{-4}$ is a threshold velocity value,
$\bar{v}(t)=(1/N)\sum_{i=1}^N v_i$ (with $v_i = \mathrm{d}x_i/\mathrm{d}t$)
is the spatially averaged dislocation velocity, and $r=100$ is a shape
parameter. In practice this amounts to a continuous, smooth approximation of
the step function, with $\mathrm{d}\sigma/\mathrm{d}t \approx m_{\sigma}$ for
$\bar{v}<v_{\mathrm{th}}$ and $\mathrm{d}\sigma/\mathrm{d}t \approx 0$ for
$\bar{v}>v_{\mathrm{th}}$, such that $\sigma$ is increased only in between
strain bursts and kept constant during them. Strain
$\epsilon$ is accumulated by dislocation motion,
\begin{equation}
\epsilon(t) = \frac{1}{N}\sum_{i=1}^{N}[x_i(t)-x_i(0)].
\end{equation}
The simulations are run until the dislocations start to flow indefinitely,
i.e., the strain burst size diverges, which happens at a sample-dependent
flow stress $\sigma_\textrm{flow}$. The stress--strain curves $\sigma(\epsilon)$ are recorded by storing
$\sigma(t)$ and $\epsilon(t)$ every $20$ dimensionless time unit during the simulations. Example stress--strain curves are shown in Fig.~\ref{fig:1}(b) for 
three different system sizes. Notice the characteristic
irregular staircase-like structure of the curves, as well as the larger
size of strain bursts for smaller systems.

The resulting stress--strain curves, each corresponding
to a different realization of the random pinning landscape and relaxed
initial dislocation positions, are stored in a database consisting of $10\,000$ stress--strain curves for each system size, to be used as training and testing data for the predictive models. For the largest systems ($N = 64\ldots 512$), we also generate a separate dataset of $1\,000$ stress--strain curves stored at a finer time resolution of $\Delta t = 2$, in order to better detect individual strain bursts. This latter data set is used to analyze the statistical properties of the strain bursts, as well as to train and test ML models for predicting strain bursts.

\section{Predictive models: from linear regression to convolutional neural networks}

In what follows we will introduce the predictive models used in this study.
These encompass both linear and non-linear models using hand-picked features as input, as well as a CNN
that requires less feature engineering as it uses a complete description of the system's initial state as input. 
% that does not require any feature engineering as it uses a complete description of the system's initial state as input. 
Using these different models allows us to assess the dependence of our results on any particular model. 
% Training of the predictive models is performed using the Keras library/interface of Python programming language. The datasets used, each consisting of 10 000 samples, are divided randomly into a training set (80\% of the data) and a testing set (the remaining 20\%, not used for training). All prediction results are results on the testing sets, averaged over 5 training instances, each with different division into training and testing sets. Loss is defined to be the mean squared error, with the addition of a possible regularization penalty (see below).

Before training, each input feature (each input channel in the CNN case) is standardized by subtracting the mean and dividing by the standard deviation, and the dataset is divided randomly into a training set containing 80\% of the data points, and a testing set containing the remaining 20\%. 
%Predictive models are trained with 80\% of the data and tested with the remaining 20\%. Before this, datasets are standardized by subtracting the mean and dividing by the standard deviation of each feature channel. 
Predictability is measured by calculating the correlation coefficient between predicted and target outputs of the testing set and averaging the result over 5 training instances. Predictive models are implemented using the Python libraries \textit{scikit-learn} \cite{scikit-learn} for LASSO and \textit{Keras} \cite{keras} for simple neural networks and CNNs.

%The prediction problems of this study are regression problems by nature, so \textit{mean square error} is selected as the main part of \textit{loss}, which is the property that a training algorithm tries to minimize. 
The prediction models work by minimizing the \emph{loss}, which mainly consists of the mean squared error. In addition, most models are $L^1$-regularized, which adds a penalty term to the loss to prevent overlearning. $L^1$ regularization often improves the result and encourages the model to focus on the most useful/promising input features while discarding the unimportant, and as a byproduct it allows us to determine which input features are used by the model.

\subsection{Linear regression: LASSO}

% The $L^1$-regularized linear regression model (LASSO, or linear absolute shrinkage and selection operator \cite{tibshirani1996regression}) is one of the two models considered here using hand-picked features (the other being simple neural network, see below). The model uses a built-in optimizer...

A linear regression model performs an affine transformation, where input features are first linearly combined by multiplication with a (weight) matrix and then biased by adding a translation vector. If the model is optimized using $L^1$-regularization for the weight matrix, it is commonly called LASSO (linear absolute shrinkage and selection operator \cite{tibshirani1996regression}). In this study, this model is applied by using \verb|sklearn.linear_model.Lasso|, which has a built-in optimizer and only requires the user to choose the regularization parameter $\alpha$. We found that values of $\alpha$ in the range $10^{-3} \lesssim \alpha \lesssim 10^{-2}$ gave good results, and chose $\alpha = 10^{-2}$.

\subsection{Simple neural networks}

% The simple neural networks considered here use {\it rectified linear unit (ReLU)} as the activation function, and {\it Adam} as the optimizer. Regularization parameter values and the numbers of hidden units for the simple neural network models are listed in Table 1...

When two linear regression models are stacked in such a way that the intermediate result is \textit{activated} by applying a non-linear activation function, such as the \textit{rectified linear unit (ReLU)}, the combined model becomes a simple neural network which can, to some extent, learn general non-linear mappings. An intermediate result of such model is called a \textit{hidden layer}, and elements of the intermediate result are referred to as \textit{hidden units}. $L^1$-regularization is applied to each layer, with a parameter $\alpha_n$ describing the strength of regularization at each layer.

Tested variants for stress--strain curve prediction differ by the number of hidden units, hidden layers and regularization. Most have one hidden layer with 64 hidden units and use 
%(\textit{kernel\_regularization}) 
$\alpha_1 = 10^{-2}$ for the first weight matrix and $\alpha_n = 10^{-5}$ for the rest. All models use ReLU-activations for the hidden layers. One variant has less regularization, meaning that $\alpha_1 = 10^{-3}$ instead of $10^{-2}$. Theoretically, one hidden layer is enough for reproducing any continuous mapping given enough hidden units, but a model with 3 hidden layers is employed here as well for comparison to other studies. The models are trained using the \textit{Adam} optimizer \cite{kingma2014adam} with learning rate $10^{-3}$ for 100 epochs.

The predictability of strain bursts is studied using a simple neural network having one hidden layer with 64 hidden units. Model output is in this case very big, consisting of a vectorized (one-dimensional) version of an avalanche map that is originally two-dimensional (see the Results section below for details). We found that all parameters collapsed to 0 if $\alpha$ is too high, so a value of $10^{-5}$ is used for all weight matrices. A lower $\alpha$ parameter makes the model susceptible to overlearning. Therefore, models are trained only for 10 epochs but with a higher learning rate of $10^{-2}$.

\subsubsection{Selecting features for prediction}

\begin{figure}[t!]
\begin{center}
  \includegraphics[width=\columnwidth]{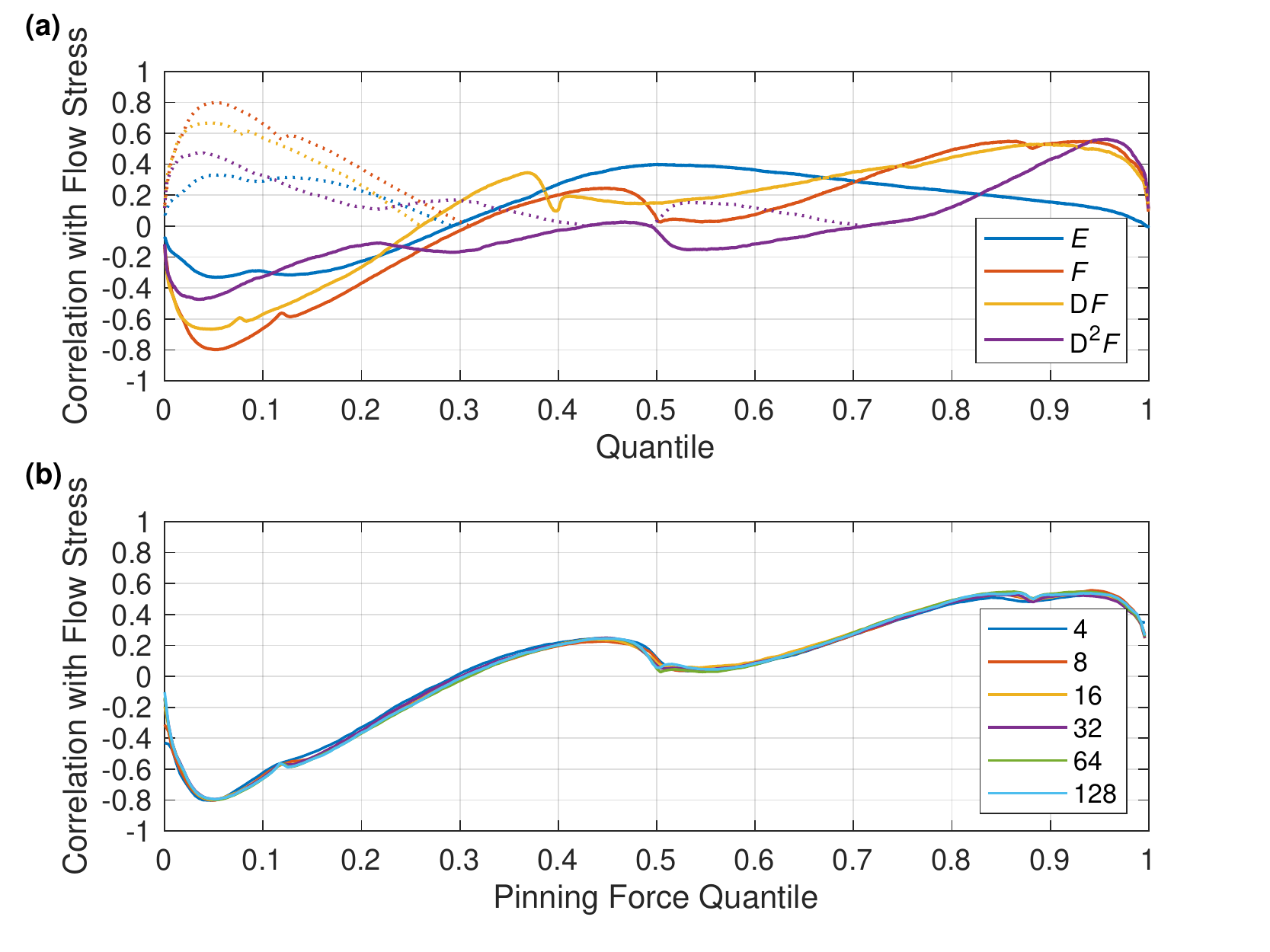}
  \caption{(a) Correlation between flow stress and quantiles of pinning potential energy $E$, pinning force $F$, derivative of the pinning force $D(F)$, and the second derivative of the pinning force $D^2(F)$ for a system of $N=64$ dislocations. Dotted lines show the absolute value of the correlation for easier comparison of magnitudes. (b) demonstrates that the correlations of the pinning force quantiles are not dependent on the system size $N$ (legend). These quantiles are used as hand-picked features for linear regression and simple neural networks. Notice that the 5\% quantile of the pinning force exhibits a high (negative) correlation of $\approx 0.8$ with the flow stress.}
  \label{fig:4}
    \end{center}    
\end{figure}

The LASSO and simple neural network models require hand-picked features for predicting stress--strain curves and strain bursts. It turns out that
good, information-rich features for flow stress prediction are given by
quantiles of distributions derived from the sample-specific pinning landscapes. Fig.~\ref{fig:4}(a) shows the correlation of the flow stress with
quantiles of the pinning energy $E$, the pinning force $F$, as well as $DF = \od{F}{x}$ and $D^2F = \od[2]{F}{x}$. 
In a non-interacting system, the flow stress would be controlled by the most negative pinning force $F$, which provides the strongest obstacle for dislocation motion. In an interacting system, it is possible for the dislocations to push on each other and overcome the strongest pinning force at a smaller applied stress. As a result, the correlation between the most negative value of $F$ [given by the $0\%$ quantile of $F$ in Fig.~\ref{fig:4}(a)] and the flow stress is quite small.
Interestingly, the 5\% quantile of $F$ exhibits a much larger negative correlation of approximately $-0.8$ with $\sigma_{\textrm{flow}}$, independently of system size [Fig.~\ref{fig:4}(b)]. Other quantiles of the pinning force as well as of the other quantities listed
in Fig.~\ref{fig:4}(a) also appear to  contain useful information for flow stress
prediction. We therefore give the entire quantile curve (sampled on 64 uniformly chosen points) of each field as input features.

In addition to these input features describing the pinning landscape, we also give input features that describe the initial relaxed configuration of dislocations $\{x_j\}_{j=1}^N$. In particular, we use the values $E(x_j), F(x_j), DF(x_j)$ and $D^2F(x_j)$, as well as the distances between successive dislocations, $Dx_j = x_{j+1} - x_j$ (with $x_{N+1} = x_1 + L$). These fields are given as sorted lists, so they can be interpreted as quantiles.

Finally, we choose a set of features given by local extrema of $E$ and $F$, as well as the successive differences in energy extrema $D(E \text{ extrema})$, also given as 64 uniformly spaced quantiles. 
%sorted lists, though interpolated so that they contain a constant amount (64) of quantile values.

\subsection{Convolutional neural networks}

These models take discretized fields as input and have the ability to find out if the target output depends on the spatial variations within the input fields. The CNN structure used in this study is adapted for periodic input signals by applying a convolution-activation-pooling scheme that continues to downsample the input array down to a single spatial element. Initially, an input field has 1\,024 = $2^{10}$ spatial elements, so there are 10 convolutional layers (each layer halves the number of spatial elements). This makes the result effectively independent of the particular choice of origin of the input fields. 
%Elements outside boundaries are fully defined in periodic systems so there is no need to worry about boundary effects.

All convolutional layers employed here use a window size of 3, periodic boundary conditions, ReLU-activation and non-overlapping max-pooling of size 2 (taking the maximum value within non-overlapping bins), and have 16 hidden units. The leftover hidden units after the last pooling operation are directly connected to the output layer. When predicting a single output feature from 2 input fields (channels), the CNN has 7\,185 learnable parameters.

The CNN is trained using the same \textit{Adam} optimizer as for other neural network models, with a learning rate of $10^{-3}$, learning rate decay of $10^{-5}$ and 50 training epochs. We found that even weak $L^1$-regularization caused the parameters to collapse to 0 in most cases, so we do not regularize these models. However, in the case of predicting flow stress only, a kernel regularization value of $5\cdot 10^{-4}$ for the convolutional layers is found to work.

The input to the CNNs consists of the pinning landscape fields $E, F, DF$ and $D^2F$ (although we found that giving one pinning field, such as $E$ or $F$, suffices because the CNN can compute the higher derivatives by differentiation), as well as a field $J$ defined as a sum of Gaussians centered on the relaxed dislocation positions $x_i$,
\begin{equation}
J(x) = \sum_{i=1}^N   \frac{1}{\sqrt{2\pi}}\mathrm{e}^{-\frac{(x-x_i)^2}{2}}.
\end{equation}
This field $J$ serves to describe the dislocation positions in terms of an input field with similar structure as the pinning landscape, and which is independent of the spatial resolution  [see Fig.~\ref{fig:7}(a) for examples of these input fields]. Notice that by contrast to the case of hand-picked features, these fields are given in order rather than sorted.

\begin{figure}[t!]
        \begin{center}
  \includegraphics[width=\columnwidth]{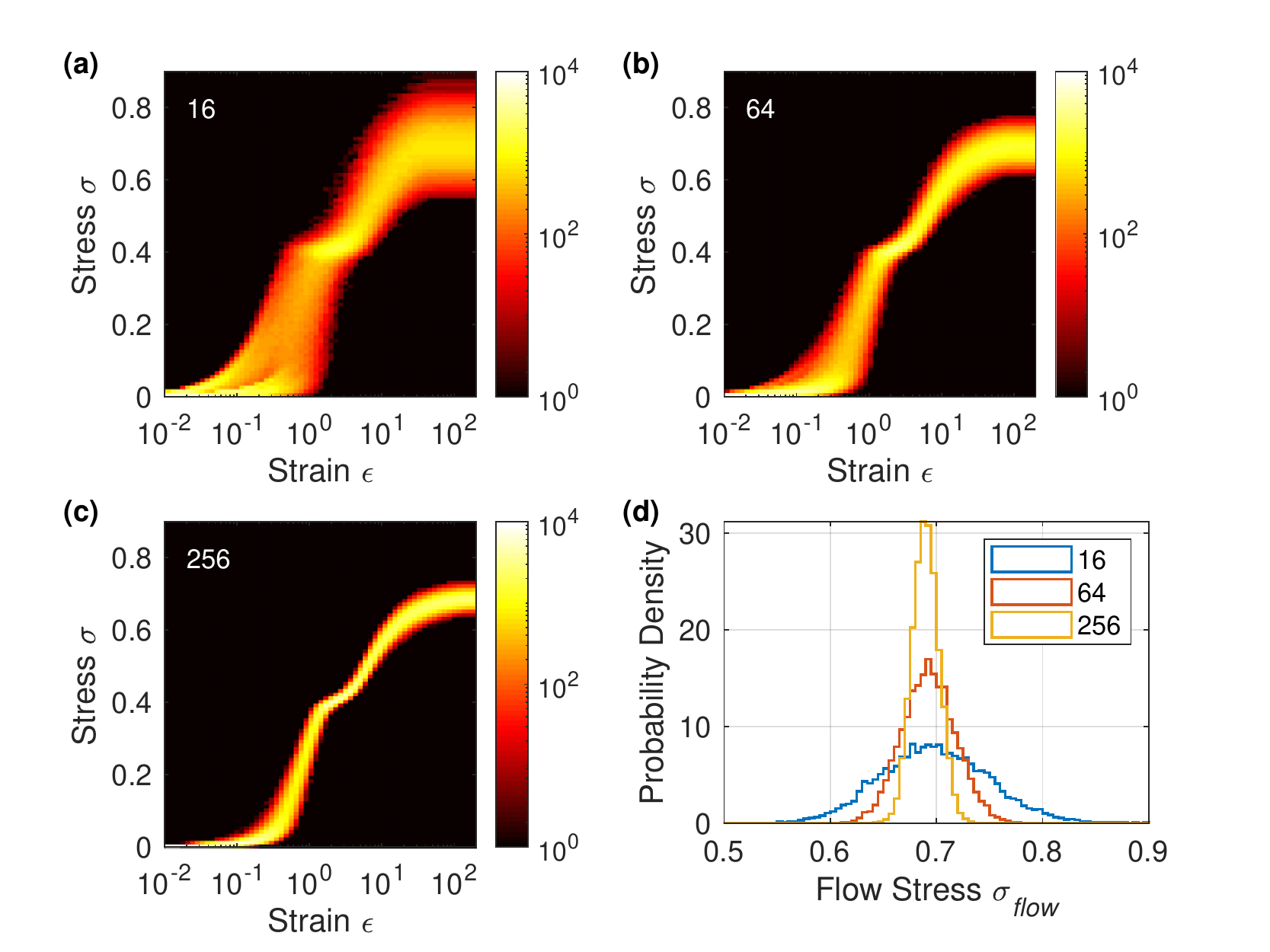}
  \caption{(a), (b) and (c) show the bivariate histograms of the stress--strain curves for different system sizes [$N=16$ in (a), $N=64$ in (b), and $N=256$ in (c)]. (d) displays the probability distributions of the sample flow stresses $\sigma_{\mathrm{flow}}$, showing how the distributions
    get increasingly narrow upon increasing the system size.}
  \label{fig:2}
      \end{center}
\end{figure}

\begin{figure}[t!]
\begin{center}
  \includegraphics[width=\columnwidth]{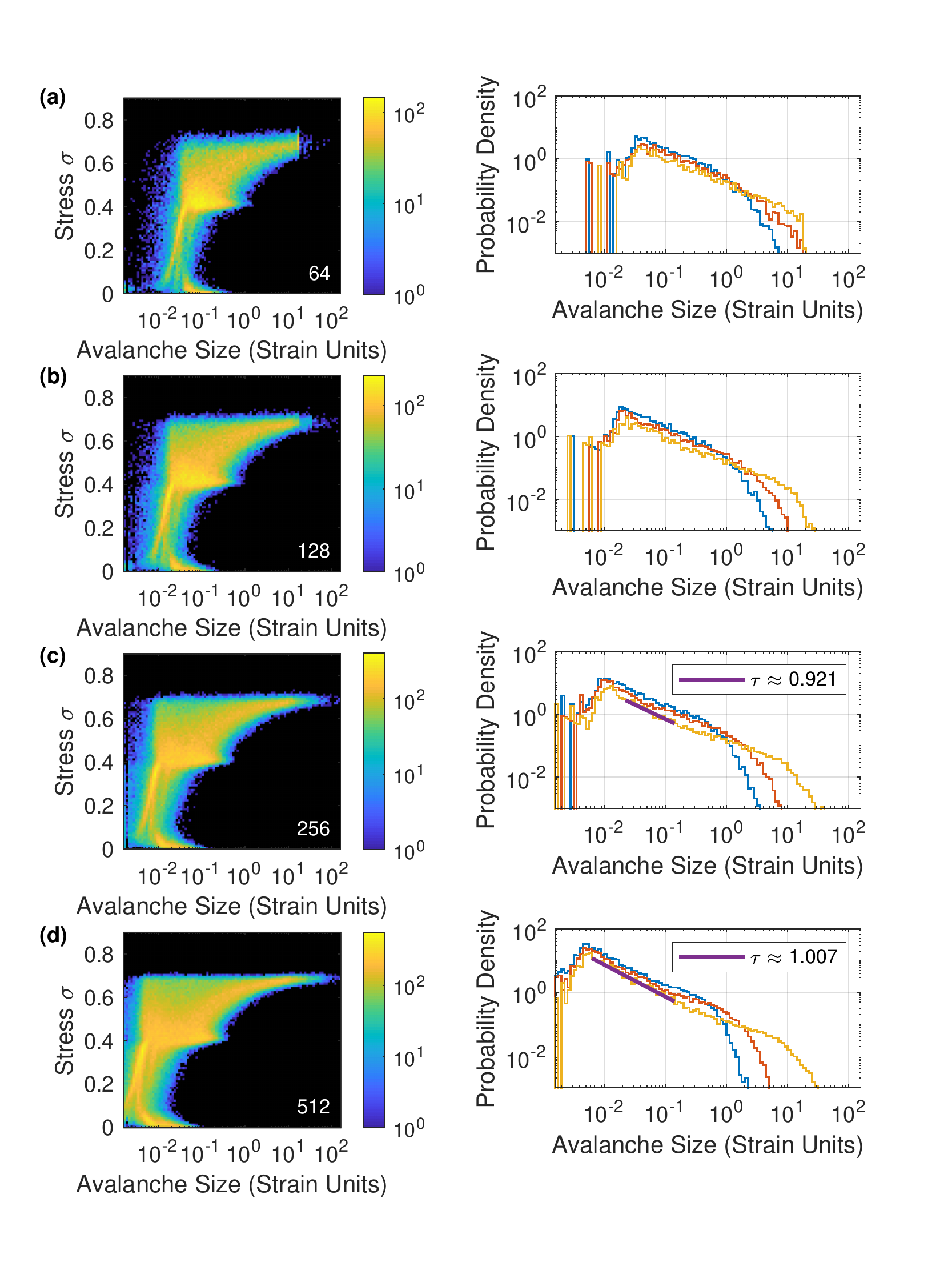}
  \caption{Left panels: Bivariate avalanche histograms (avalanche counts) as a function of the stress $\sigma$ and avalanche size $\Delta \epsilon$. Right
    panels: The stress-resolved probability distributions of the strain bursts sizes $P(\Delta \epsilon,\sigma)$, considering three stress bins: 0.59-0.60 (blue), 0.63-0.64 (red), and 0.67-0.68 (yellow). The average flow stress is around 0.69. (a) $N=64$, (b) $N=128$, (c) $N=256$ and (d) $N=512$. The right panel of (d) shows that the highest stress bin data in the largest system size exhibits a power-law part consistent with an exponent $\tau \approx 1.0$ [see Eq. (\ref{eq:ps})].}
  \label{fig:3}
  \end{center}
\end{figure}

\begin{figure}[t!]
\begin{center}
  \includegraphics[width=\columnwidth]{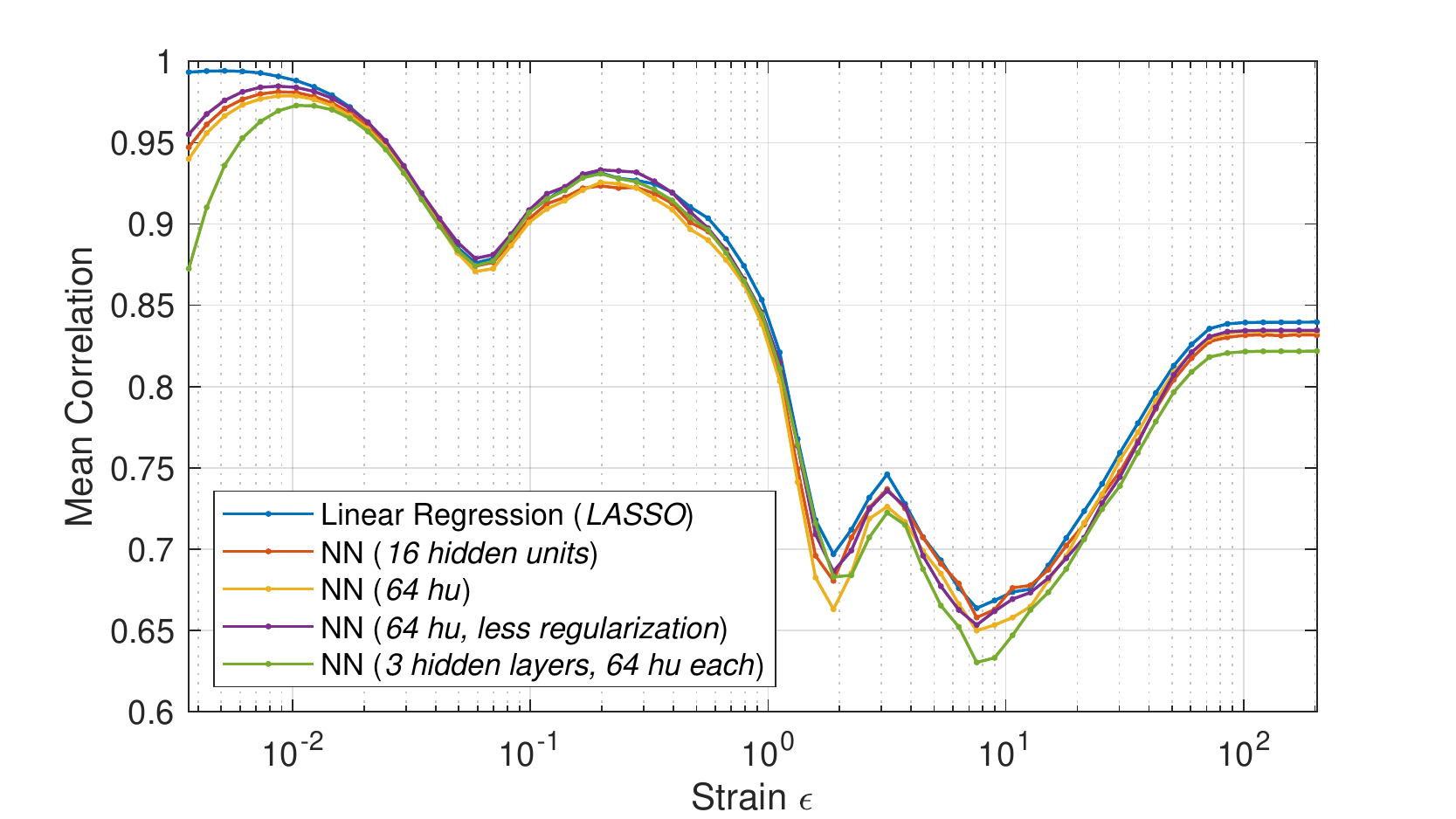}
  \caption{Predictability of stress--strain curves for the system size $N=64$ using
    hand-picked features,
    measured by the correlation between predicted and actual stresses in the test set. The shown result is a sample mean over 5 trained models, each using
    their own random division into training and testing sets. Each line
    corresponds to a given predictive model as indicated in the caption. The plateau at large strains corresponds to the predictability of the flow stress.}
  \label{fig:5}
  \end{center}
\end{figure}

\iffalse %%%%%%%%%%%%%%

\begin{table}[t]
  \centering
\begin{tabular}{ |p{6cm}||p{3cm}|p{1.5cm}|p{1.5cm}|p{1.5cm}|p{1.5cm}|  }
\hline
% \multicolumn{4}{|c|}{Country List} \\
% \hline
 Model & Operator & $L^1(M_1)$ & $L^1(M_2)$ & $L^1(M_3)$ & $L^1(M_4)$\\
 \hline
 \hline
 Linear regression (LASSO)   & $L=BM$ & $1\cdot 10^{-2}$ & - & - & -\\
 Neural network (16 hidden units)   & $LN=BM_2AM_1$ & $1\cdot 10^{-2}$ & $1\cdot 10^{-5} $& - & -\\
 Neural network (64 hidden units)   & $LN$ & $1\cdot 10^{-2}$ & $1\cdot 10^{-5}$ & - & -\\
 Neural network (64 hidden units) v2  & $LN$ & $1\cdot 10^{-3}$ & $1\cdot 10^{-5}$ & - & -\\
 Neural network ($3\times 64$ hidden units)   & $LN_3N_2N_1$ & $1\cdot 10^{-2}$ & $1\cdot 10^{-5}$ & $1\cdot 10^{-5}$ & 1$\cdot 10^{-5}$\\
 \hline
\end{tabular}
\caption{Parameters of the LASSO model and different simple neural networks used. The number of hidden units is the number of features (here elements, generally channels) in the intermediate result (change in dimensionality is a consequence of matrix operators $M_i$). The fifth model has 3 intermediate results (hidden layers) having 64 hidden units each. The second column shows the equivalent model operator using matrix formalism (see text). Columns 3-6 show the $L_1$ penalty parameter applied to each weight matrix of the models.}
\label{tab:1}
\end{table}

\fi %%%%%%%%%%%%%

\section{Results}

In this section, we first consider the statistical properties of our dataset of
stress--strain curves. Then we report our results for training the predictive models, and employ them to explore the predictability of the deformation
dynamics of the pileup model. Finally, we also discuss our results on 
the problem of predicting individual strain bursts. 

\subsection{Statistical properties of the stress--strain curves}

We start by quantifying the system size dependent sample-to-sample variations
in the plastic response of the system. Figs.~\ref{fig:2}(a-c) show bivariate
histograms of the stress--strain curves for three system sizes (with $N=16$, 64
and 256, respectively), illustrating that the sample-to-sample variability
in the plastic response is more pronounced in smaller systems. The same
applies to the probability densities of the sample flow stresses
$\sigma_\textrm{flow}$, shown in Fig.~\ref{fig:2}(d): The distributions are
quite broad for small systems and get narrower upon increasing the system
size, presumably approaching a delta peak in the thermodynamic limit. The
average magnitude of the $\sigma_\textrm{flow}$-values is in agreement with
previous results. \cite{moretti2004depinning}

\begin{figure}[t!]
\begin{center}
  \includegraphics[width=\columnwidth]{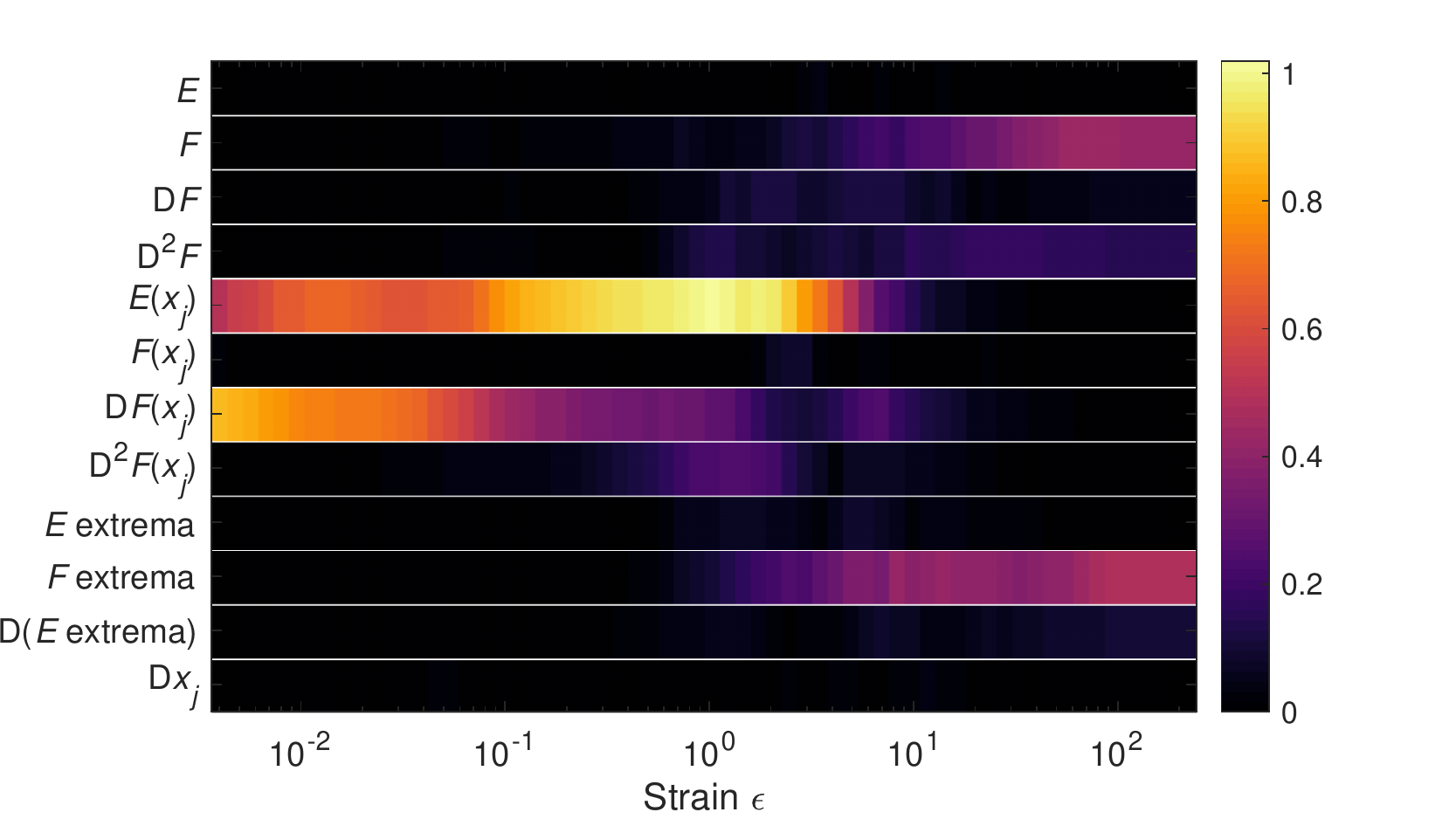}
  \caption{Weight analysis of the trained LASSO model for the $N=64$ system. Color indicates the total $L^1$ penalty ($L^1$ norm) from each feature category’s weights as a function of strain, averaged over 5 training instances. Notice that properties of the initial dislocation positions are most useful at small strains, while more generic features of the force landscape become more useful at larger strains.
  %Categories 1-4 (starting from the upmost) contain quantiles of the pinning energy, force, force derivative and force second derivative distributions. Categories 5-8 are quantiles of distributions of the same quantities as 1-4 but evaluated only at the relaxed dislocation positions (at the beginning of the stress--strain curve simulation). Categories 9-11 are quantiles from distributions derived from local extrema of the pinning fields, and category 12 contains the quantiles of the distances between neighboring relaxed dislocations. $x_j$ refers to the relaxed dislocation positions.
  }
  \label{fig:6}
\end{center}
\end{figure}

As the stress--strain curves consist of a sequence of discrete strain bursts
or dislocation avalanches $\Delta \epsilon$, we consider also their
stress-dependent size distributions $P (\Delta \epsilon,\sigma)$. Quite
generally for systems exhibiting a
depinning phase transition, one expects them to obey the scaling form
\begin{equation}
\label{eq:ps}
P(\Delta \epsilon,\sigma) = (\Delta \epsilon)^{-\tau} f\left(\frac{\Delta \epsilon}
{\Delta \epsilon^*(\sigma)}\right),
\end{equation}
where $\tau$ is the critical exponent of the avalanche size distribution,
$f(x)$ is a scaling function that obeys $f(x)=\mathrm{const.}$ for $x \ll 1$
and $f(x) \rightarrow 0$ for $x \gg 1$, and $\Delta \epsilon^*(\sigma)$ is a
stress-dependent cutoff avalanche size. Left panels of Figs.~\ref{fig:3}(a-d)
show the bivariate histograms of the avalanche count as a function of $\sigma$
and avalanche size $\Delta \epsilon$ for different system sizes (with
$N=64$, 128, 256 and 512, respectively). These illustrate that indeed the
maximum strain burst size increases upon approaching the flow stress from
below, and the increase close to the flow stress is more clear-cut for
larger systems. Notice also that these histograms show an excess of strain bursts around $\sigma \approx 0.4$. This feature is likely related to the characteristic pinning force magnitude due to an isolated pinning point given by Eq. (\ref{eq:pf}). Right panels of Figs.~\ref{fig:3}(a-d) show the corresponding
stress-resolved strain burst size distributions $P(\Delta \epsilon,\sigma)$,
considering three different stress bins below the flow stress for each system
size. For the largest system sizes considered, the data appears to be
consistent with a power-law part characterized by an exponent $\tau \approx 1.0$ [see right panels of Figs.~\ref{fig:3}(c) and (d)]. This value is somewhat lower than the expectation of $\tau \approx 1.25$ based on the scaling of the elastic energy of the pileup in Fourier space, \cite{moretti2004depinning} suggesting that the pileup model would be in the same universality class with depinning of contact lines \cite{joanny1984model} and planar cracks. \cite{bonamy2008crackling,laurson2010avalanches} As we focus here on studying rather small systems, this difference in likely due to finite size effects.

%Interestingly, this exponent value deviates from that predicted for mean field depinning ($\tau_\textrm{MF}=1.5$, see, e.g., Ref. \cite{zapperi1998dynamics}) even if dislocations of the pileup interact via long-range stress fields. Note that similar deviations from mean field scaling have been observed before: The critical relaxation of the order parameter (strain rate/mean dislocation velocity) has been found to decay in time as $t^{-\theta}$, with $\theta \approx 0.65$, \cite{moretti2004depinning} while mean field theory would result in $\theta=1$. 

Overall, the statistical properties of the plastic response of the pileup
model shows two key features relevant for deformation predictability: (i)
There are sample-to-sample variations in the response (that we will try to predict using machine learning), and (ii) the distribution of strain bursts $\Delta \epsilon$
displays characteristics of a critical system, suggesting that individual
strain bursts may be inherently unpredictable. In what follows we will
use our trained predictive models to explore how well (i) the sample
stress--strain curves and (ii) individual strain bursts can be predicted.

\begin{figure}[t!]
\begin{center}
  \includegraphics[width=\columnwidth]{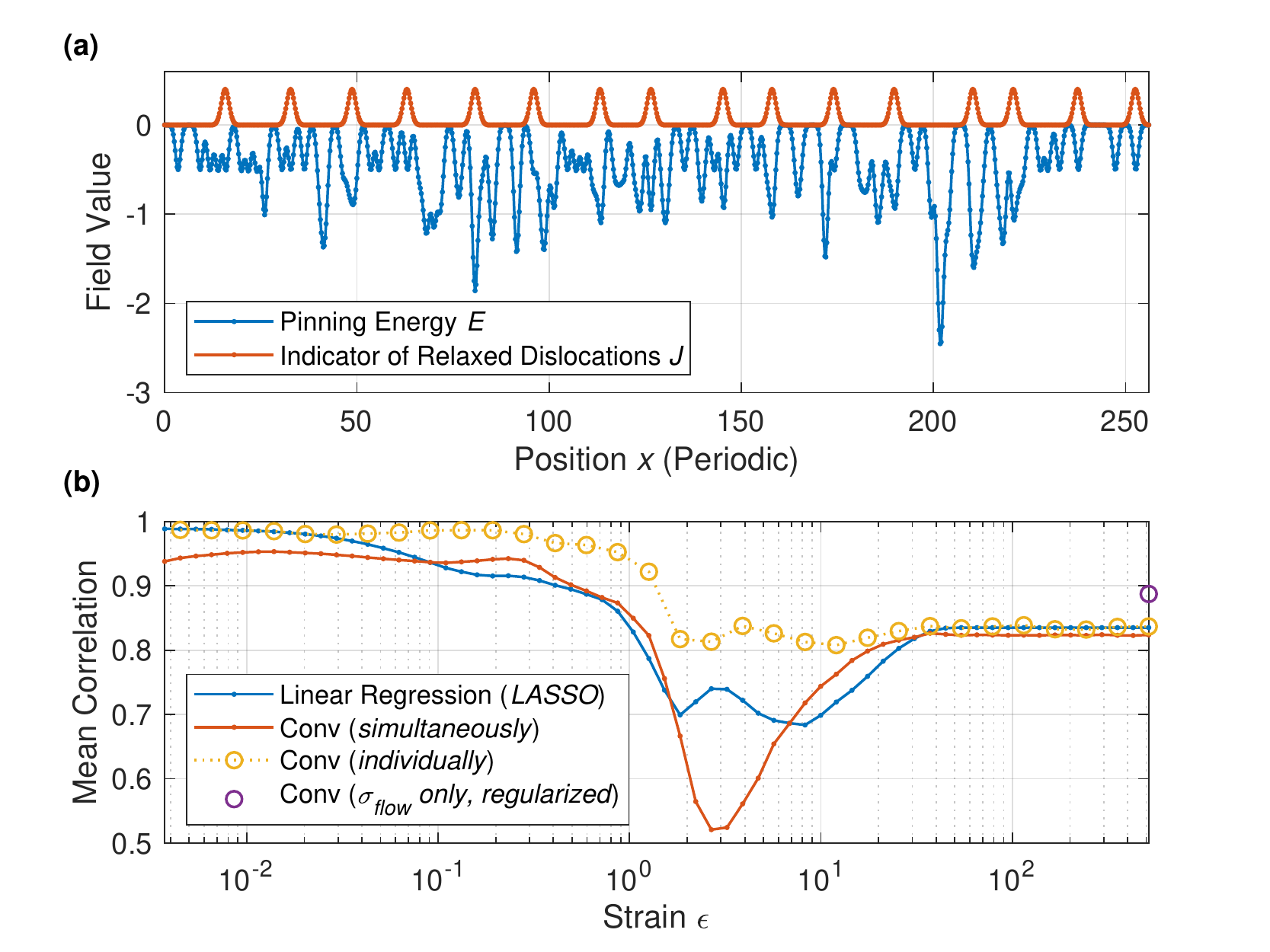}
  \caption{(a) Example input fields for a CNN with $N=16$. Dot markers show the spatial resolution. (b) Stress--strain curve predictability at $N=16$ for different models: LASSO (blue line); a "generalist" CNN predicting the entire stress--strain curve at once (red line); individual "specialist" CNNs trained specifically to predict the stress at a single strain (yellow circles); and an $L^1$-regularized CNN trained to only predict the flow stress (purple circle).}
    \label{fig:7}
    \end{center}
\end{figure}

\subsection{Predictability of the sample stress--strain curves}

\subsubsection{LASSO and simple neural networks}

Fig.~\ref{fig:5} shows the mean correlation coefficient between the predicted and actual stress values as a function of strain for $N=64$, 
using linear regression (LASSO) and different variants of simple neural networks.
%; the models used along with the $L_1$ regularisation parameters employed are listed in Table \ref{tab:1}. 
First, we note that the flow stress prediction is only slightly better than the magnitude of the correlation with the 5\% quantile of the pinning force landscape found above ($\sim 0.8$, see Fig. \ref{fig:4}), with the correlation coefficient assuming values in the range of 0.825-0.84 depending on the model used (the large-strain plateau in Fig.~\ref{fig:5}). Second, the correlation clearly depends on strain, exhibiting a rather pronounced dip reaching values as low as 0.65 or so for intermediate strains, along with two additional, lesser dips for smaller strains, before reaching a value close to 1 as the strain goes to zero. The latter is true especially for LASSO, while the neural network models seem to struggle a bit in capturing the deformation dynamics for very small strains. The reduced predictability for intermediate strains could be related to the importance of inherently unpredictable dislocation avalanches, in analogy with results of Ref. \cite{salmenjoki2018machine}.

It is interesting to analyze which features the predictive models end up using for prediction, and how that might depend on strain. Since the models utilize heavy $L^1$-regularization, analysis of the learned weights becomes straightforward. Features that contribute to predictability are allowed to have high regularization penalty in exchange for decreasing loss, while unimportant features will not decrease loss sufficiently to justify a high regularization penalty. Fig.~\ref{fig:6} shows how much penalty is generated by the weights corresponding to each feature category for the LASSO model. At the start, the pinning force derivative at the relaxed dislocation positions, $DF(x_j)$, is the most active feature, but also the pinning energy at the initial dislocation positions, $E(x_j)$, is used. This is reasonable as such quantities could be used to estimate the initial linear response of the dislocations. Interestingly, and perhaps somewhat counter-intuitively, the pinning energy at the initial dislocation positions becomes even more important for intermediate strains, in addition to a larger diversity of features used. Flow stress prediction uses mainly the quantiles of the pinning force $F$, as well as its local extrema points. This dependence of the important features on strain indicates that predicting the response of the system to infinitesimal stresses is of a different nature compared to predicting the flow stress.

\subsubsection{Predictability of stress--strain curves: convolutional neural networks}

We then proceed to employ CNNs as predictive models. Fig.~\ref{fig:7}(b) shows the mean correlation as a function of strain as obtained using three different types of CNNs, along with the LASSO result for reference. Here, the system size $N=16$ is considered. 
First, we consider a CNN which predicts the entire stress--strain curve at once [a "generalist" CNN, shown as a red line in Fig.~\ref{fig:7}(b)]. This model actually underperforms the simple linear regression model for majority of the strain values. This could be due to the fundamentally different nature of the problem for different strain values, as illustrated by Fig.~\ref{fig:6}, showing that different features are important for different strains. Hence, the "generalist" CNN might get exhausted and not be able to optimally learn the entire stress--strain curve at once.

Thus, we consider also a "specialist" CNN which is trained separately for each strain value to learn the corresponding value of stress. The result is shown as yellow circles in Fig.~\ref{fig:7}(b). Indeed, as compared to the generalist model, the specialist model is able to learn the deformation dynamics of the pileup model much better, especially for intermediate strains, where the generalist CNN exhibits a deep dip in the correlation coefficient, presumably due to the importance of hard to predict dislocation avalanches for those strains. It is interesting that the specialist model is able to reach a correlation coefficient exceeding 0.8 also in this regime. 

Notably, neither of the models discussed above is able to predict the flow stress much better than by just considering the correlation with the 5\% quantile of the pinning force $F$ (see Fig.~\ref{fig:4}). We therefore try to further improve the specialist CNN by switching on $L^1$-regularization and using only the pinning force $F$ as the input field (this was shown above to be the most informative quantity for flow stress prediction). The individual purple point at a large strain value in Fig. \ref{fig:7}(b) is the resulting flow stress prediction, with a correlation coefficient of 0.89, i.e., significantly better correlation than using any of the other models considered.

\subsection{Predictability of strain bursts}

\begin{figure}[t!]
\begin{center}
  \includegraphics[width=\columnwidth]{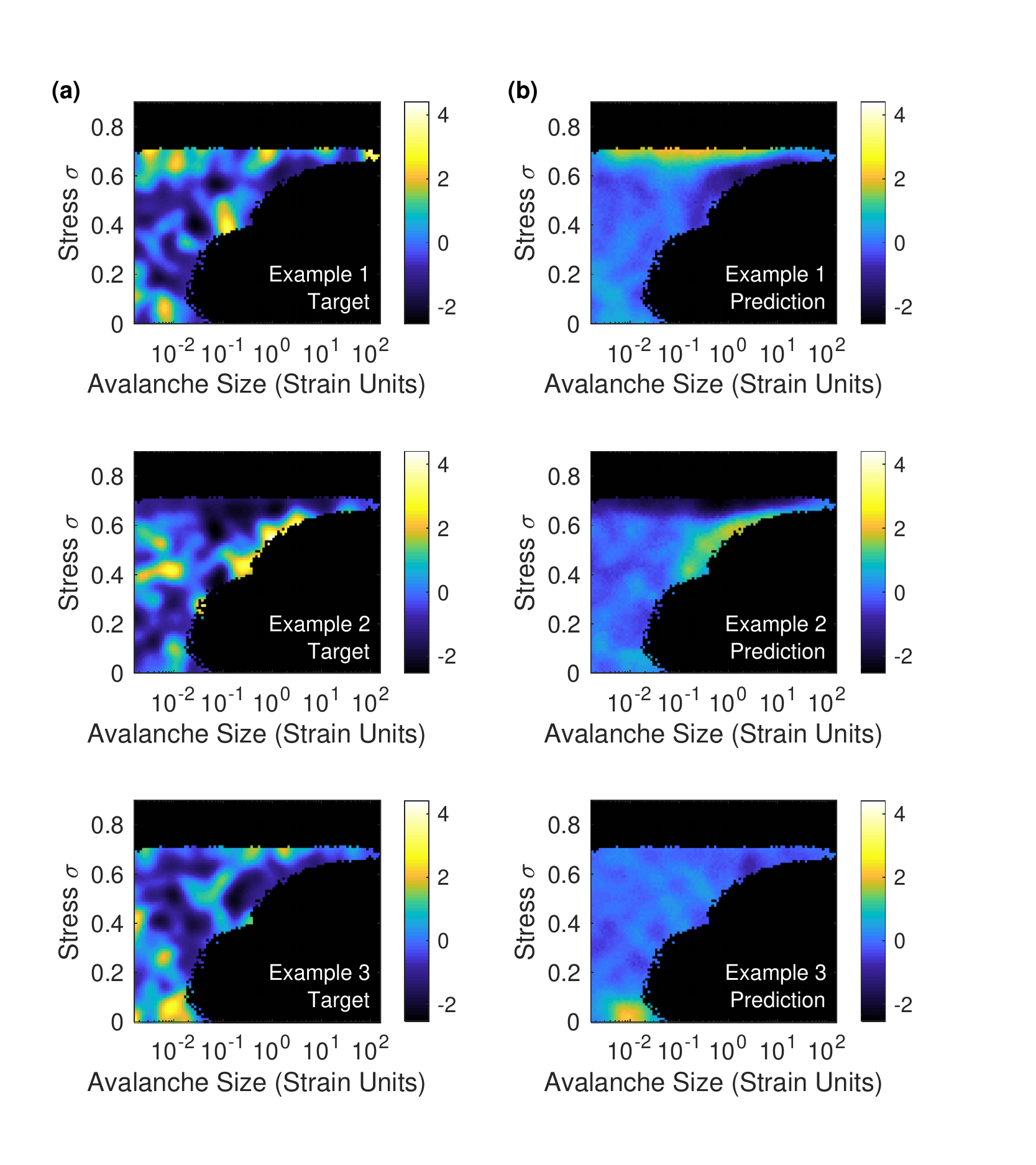}
  \caption{Three examples (rows) of strain burst maps of individual systems for $N$=512. Left column (a) shows 3 target maps, and on the right (b) are the corresponding predictions of a simple neural network.}
    \label{fig:8}   
    \end{center}
\end{figure}

Finally, we explore the possibility of predicting individual
strain bursts along the stress--strain curves. First, we need to formulate a problem that is suitable for analysis using the predictive models at hand. To this end, we generalize the bivariate avalanche histograms (left panels of Fig.~\ref{fig:3}) to the case of individual samples. As such, these would be rather sparse sets of points in the space spanned by avalanche size and stress. We therefore smooth out these maps by convolving with a Gaussian with a standard deviation of 0.03 units of stress in the vertical direction, and of 0.15 units of $\log_{10}(\text{avalanche size})$ in the horizontal direction; both correspond to 3 times the bin size used in Figs.~\ref{fig:3} and \ref{fig:8}. The studied region is limited to bins where at least one avalanche appears within the dataset, so that the Gaussian smoothing does not cause distribution boundaries to extend to impossible areas. Lastly, the maps are standardized by removing the mean and dividing by the standard deviation at each bin. We then employ a simple neural network with 64 hidden units to find a mapping from the hand-picked input features to the smoothed sample-specific bivariate avalanche histograms. 

\begin{figure}[t!]
\begin{center}
  \includegraphics[width=\columnwidth]{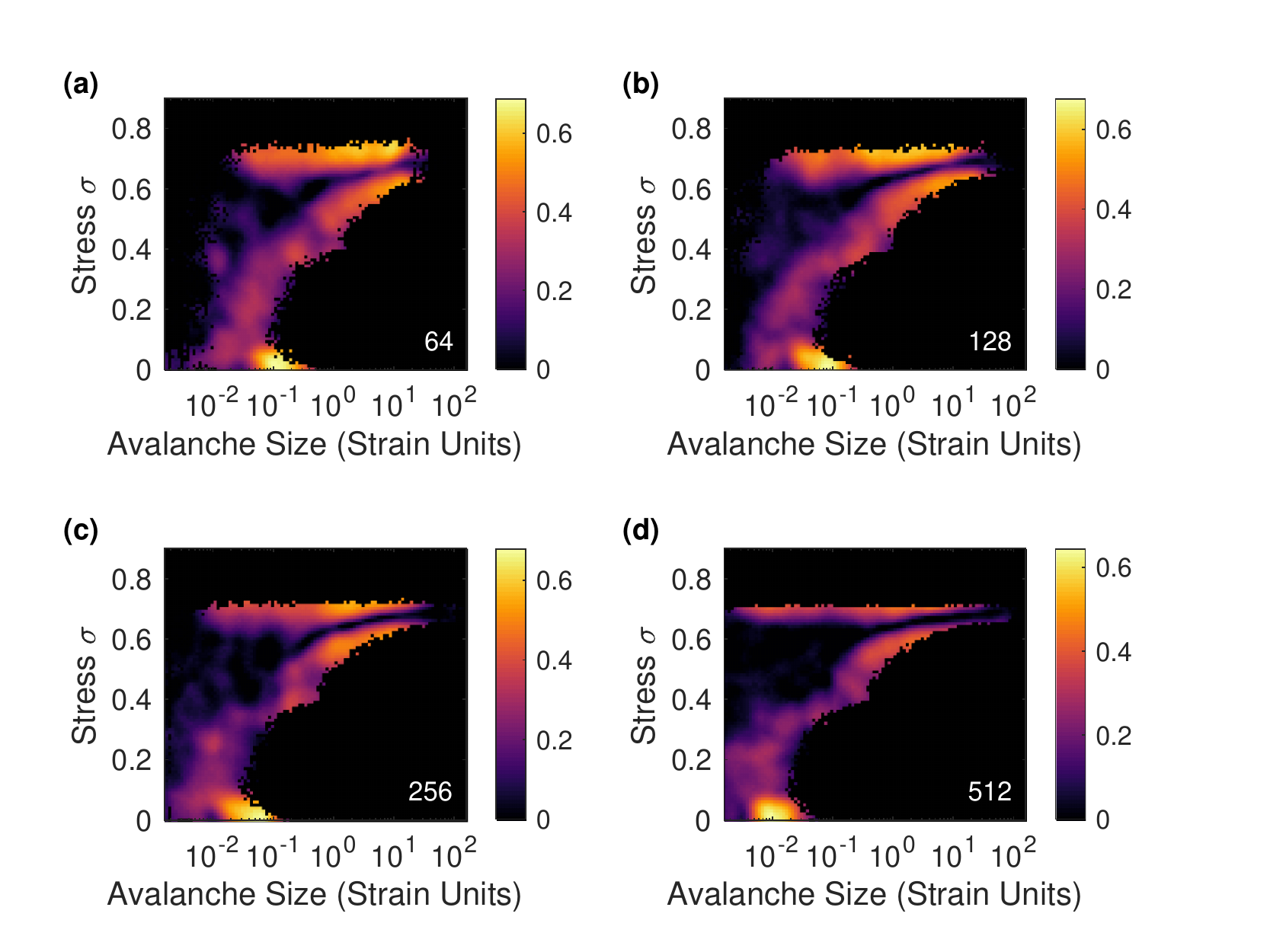}
  \caption{Correlation maps between the predicted and actual
    sample-specific avalanche maps for different system sizes, (a) $N=64$, (b) $N=128$, (c) $N=256$ and (d) $N=512$. Notice the positive correlations for the largest avalanches for each stress value (avalanche cut-off) and for most of the avalanches very close to the flow stress, as well as the approximately zero correlation for avalanches belonging to the power law scaling regime (see Fig.~\ref{fig:3}).}
  \label{fig:9}
  \end{center}
\end{figure}

Three different examples of smoothed avalanche maps are shown in the left column of Fig.~\ref{fig:8}, with corresponding predicted avalanche maps in the right column. While the predictions are clearly not perfect, the algorithm is able to capture some key features of these avalanche maps.
Fig.~\ref{fig:9} shows correlation maps between the predicted and actual avalanche densities at 4 different system sizes, averaged over 5 training instances. These show a number of interesting features: (i) Large parts of the sample-specific avalanche histograms appear to be completely unpredictable, with the correlation coefficient assuming a value very close to zero (shown in black in Fig.~\ref{fig:9}). This is especially true for most of the regions corresponding to the power-law scaling regime of the avalanche size distribution, in agreement with the idea that critical avalanches should be intrinsically hard to predict. \cite{pun2020prediction} (ii) Avalanches with the largest size for each stress value (i.e., those belonging to the cutoff of the stress-resolved avalanche size distribution) appear to exhibit some degree of predictability, although the related correlation coefficient values are somewhat lower than those found for predictability of the stress--strain curves. This is agreement with the idea that avalanches in the cutoff are not critical and hence somewhat predictable. (iii) Surprisingly, we also find non-vanishing predictability of avalanches of any size occurring in the immediate proximity of the flow stress (bright horizontal segments in the upper parts of the panels of Fig.~\ref{fig:9}). This is surprising because those events are expected to be critical and hence should be unpredictable. 

\begin{figure}[t!]
\begin{center}
  \includegraphics[width=\columnwidth]{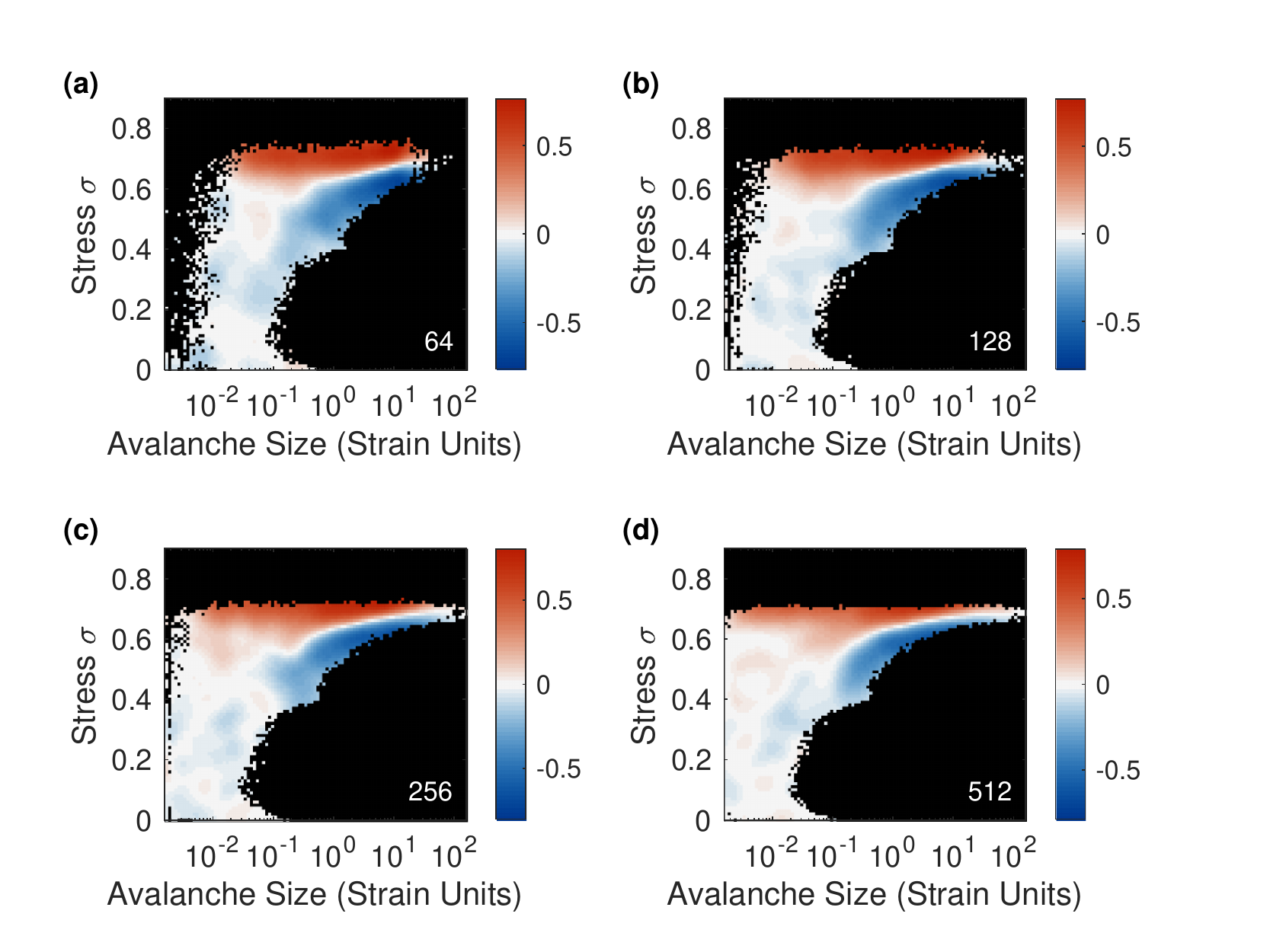}
  \caption{Correlation maps between the avalanche count and flow stress for the avalanche maps considered in Fig.~\ref{fig:8}. (a) $N=64$, (b) $N=128$, (c) $N=256$ and (d) $N=512$. Notice the positive correlation between the flow stress and avalanches close to the flow stress, as well as the negative correlation between the flow stress and the largest avalanches occurring for smaller stresses.}
  \label{fig:10}
\end{center}
\end{figure}

In order to shed some light on the origin of the key features of avalanche predictability discussed above, we finish by considering the correlation between the avalanche count and flow stress. The logic here is that since we have demonstrated above that the sample-specific flow stress can be predicted quite well, if the avalanche maps are correlated with the flow stress, these should be predictable as well. Fig.~\ref{fig:10} reveals two stand-out features: (i) The avalanches taking place in the immediate proximity of the flow stress exhibit clear positive correlation with the flow stress, and (ii) the largest avalanches for stresses smaller than the flow stress (i.e., avalanches belonging to the stress-dependent cut-off of the avalanche size distribution) are negatively correlated with the flow stress. This means that both a {\it higher} than average number of avalanches very close to the flow stress and a {\it lower} than average number of large avalanches at smaller stresses imply a higher than average flow stress value. The good level of predictability of the flow stress demonstrated above thus translates into reasonable predictability of these features of the sample-specific bivariate avalanche histograms.
Notice that the widths of the predictable bands in the avalanche maps (Fig.~\ref{fig:9}) appear to become thinner with increasing system size, so we cannot exclude the possibility that this avalanche predictability would be a finite size effect. 

\section{Discussion and conclusions}

We have established that predictive models ranging from linear regression to CNNs can be used to predict the stress--strain curves of the pileup model with a good accuracy. While the different models employed give rise to somewhat different results, a practical conclusion is that the model giving the highest correlation coefficient for a given strain provides a lower limit for deformation predictability of the pileup system. Notably the "specialist" CNN discussed above results in correlation coefficients exceeding 0.8 over the entire range of strains considered, and adding regularization further improves the correlation coefficient of the flow stress prediction to 0.89. These should be interpreted as lower limits of deformation predictability as we obviously cannot exclude the possibility that a hypothetical predictive model not considered here might outperform our models.

While the predictability of the stress--strain curves turns out to be quite good, it is not perfect. This is to be expected because the stress--strain curves consist of critical-like strain bursts which are expected to be inherently hard to predict. Indeed, our attempts to predict the sample-level bivariate avalanche count resulted in essentially zero predictability for most of the "critical" dislocation avalanches belonging to the power-law distributed part of the avalanche size distribution. On the other hand, these sample-level avalanche distributions also showed an interesting regularity: A larger than average flow stress value indicates lower than average number of the largest avalanches for stresses below the flow stress, and larger than average number of avalanches in the immediate vicinity of the flow stress. It would be interesting to extend such analysis from our stress-controlled simulations to strain-controlled ones. 

The pileup model studied here has the advantage that the dislocation system will sample the entire one-dimensional random energy landscape, and hence extracting relevant descriptors of the sample-specific quenched disorder is straightforward.
This is in contrast to other systems exhibiting a depinning phase transition such as magnetic domain walls in disordered ferromagnetic thin films, \cite{zapperi1998dynamics} planar crack fronts in disordered solids, \cite{bonamy2008crackling,laurson2010avalanches,laurson2013evolution} or individual dislocation lines interacting with point defects. \cite{zapperi2001depinning} In such systems, driven by an external
force in a direction perpendicular to the average elastic line direction, the disorder sampled by a given realization is not known {\it a priori}. Nevertheless, it would be interesting to extend the present study to consider the predictability of depinning dynamics in such systems. Finally, we point out two crucial outstanding questions related to deformation predictability: The problem remains to be addressed in realistic three-dimensional DDD simulations, with \cite{salmenjoki2020precipitate,lehtinen2018effects} or without \cite{lehtinen2016glassy} additional defects interfering with dislocation motion, as well as in experiments where descriptors of the initial sample microstructure could be extracted, e.g., by X-ray measurement techniques. \cite{ludwig2001three,levine2006x,schafler2005second}\\

\begin{acknowledgments}
The authors wish to thank Henri Salmenjoki for technical help on
machine learning. We acknowledge the financial support of the Academy 
of Finland via the Academy Project COPLAST (Project no. 322405).
\end{acknowledgments}
  
\section*{Data availability}  
The data that support the findings of this study are available from the corresponding author upon reasonable request.

\section*{References} 
\bibliography{aipsamp}
\end{document}